\documentclass{iaus}
\usepackage{graphics}

  \checkfont{eurm10}
  \iffontfound
    \IfFileExists{upmath.sty}
      {\typeout{^^JFound AMS Euler Roman fonts on the system,
                   using the 'upmath' package.^^J}%
       \usepackage{upmath}}
      {\typeout{^^JFound AMS Euler Roman fonts on the system, but you
                   dont seem to have the}%
       \typeout{'upmath' package installed. iaus.cls can take advantage
                 of these fonts,^^Jif you use 'upmath' package.^^J}%
      }
  \else
  \fi

  \checkfont{msam10}
  \iffontfound
    \IfFileExists{amssymb.sty}
      {\typeout{^^JFound AMS Symbol fonts on the system, using the
                'amssymb' package.^^J}%
       \usepackage{amssymb}%

      }{}
  \fi

  \IfFileExists{amsbsy.sty}
    {\typeout{^^JFound the 'amsbsy' package on the system, using it.^^J}%
     \usepackage{amsbsy}}
    {\providecommand\boldsymbol[1]{\mbox{\boldmath $##1$}}}

\newsavebox{\astrutbox}
\sbox{\astrutbox}{\rule[-5pt]{0pt}{20pt}}

\title[The Interplay among Black Holes, Stars and ISM in Galactic 
       Nuclei]{Observational Signature of Tidal Disruption of a Star\\
 by a Massive Black Hole}

\author[T. Bogdanovi\'c {\it et al.\/}]%
{Tamara Bogdanovi\'c$^1$,
Michael Eracleous, Suvrath Mahadevan,\break Steinn Sigurdsson$^1$,
and Pablo Laguna$^1$}

\affiliation{Department of Astronomy \& Astrophysics, 
The Pennsylvania State University, University Park, PA 16802, USA 
email: tamarab, mce, suvrath, steinn, pablo@astro.psu.edu\\[\affilskip]
$^1$also member of Center for Gravitational Wave Physics}

\pubyear{2004}
\volume{222}
\pagerange{1--2}
\date{?? and in revised form ??}
\jname{The Interplay among Black Holes, Stars and ISM \\in Galactic Nuclei}
\editors{Th. Storchi Bergmann, L.C. Ho \& H.R. Schmitt, eds.}
\begin{document}

\maketitle

\begin{abstract}
We have modeled the time-variable profiles of the H$\alpha$ emission
line from the non-axisymmetric disk and debris tail created in the
tidal disruption of a solar-type star by a $10^{6}\,M_{\odot}$ black
hole. We find that the line profiles at these very early stages of the
evolution of the post-disruption debris do not resemble the double
peaked profiles expected from a rotating disk since the debris has not
yet settled into such a stable structure. The predicted line profiles
vary on fairly short time scales (of order hours to days). As a result
of the uneven distribution of the debris and the existence of a
``tidal tail'' (the stream of returning debris), the line profiles
depend sensitively on the orientation of the tail relative to the line
of sight. Given the illuminating UV/X-ray light curve, we also model
the H$\alpha$ light curve from the debris.
\end{abstract}

\firstsection 
\section*{Light Curves and Emission Line Profiles From the Tidal Debris}

\par{Simulations of tidal disruption of a star were carried out using a 
three-dimensional, relativistic, smooth-particle hydrodynamics code
(\cite[Laguna et al. 1993]{Laguna}), to describe the early evolution
of the debris during the first fifty to ninety days. We have used the
photoionization code CLOUDY (\cite[Ferland 1996]{Ferland}) to
calculate the physical conditions and radiative processes in the
debris. To obtain the observed profile from the relativistic debris,
confined to a plane, in the weak field approximation we have followed
the line profile calculations by \cite[Chen \& Halpern (1989)]{CH} and
\cite[Eracleous et al. (1995)]{ELHS}.}

\par{Once bound debris starts to rain down on the black hole 
it is expected to cause the initial rapid rise in the emitted UV/X-ray
light curve and steady decay with the power low index of $-5/3$ later
on (\cite[Rees 1988]{Rees}). The UV/X-ray flares from the central
source illuminate the debris, the photons get absorbed, and some are
re-emitted in the Balmer series H$\alpha$ line. We have modeled the
H$\alpha$ light curve and emission line profiles in the period
immediately after the accretion rate onto the black hole became
significant. The main features of the H$\alpha$ light curve are an
initial rise followed by a decline, with superposed fluctuations. The
decay rate of the H$\alpha$ light curve is determined by the decay
rate of UV/X-ray light curve, and the debris expansion and
redistribution rate.}

\par
\begin{figure}
 \includegraphics{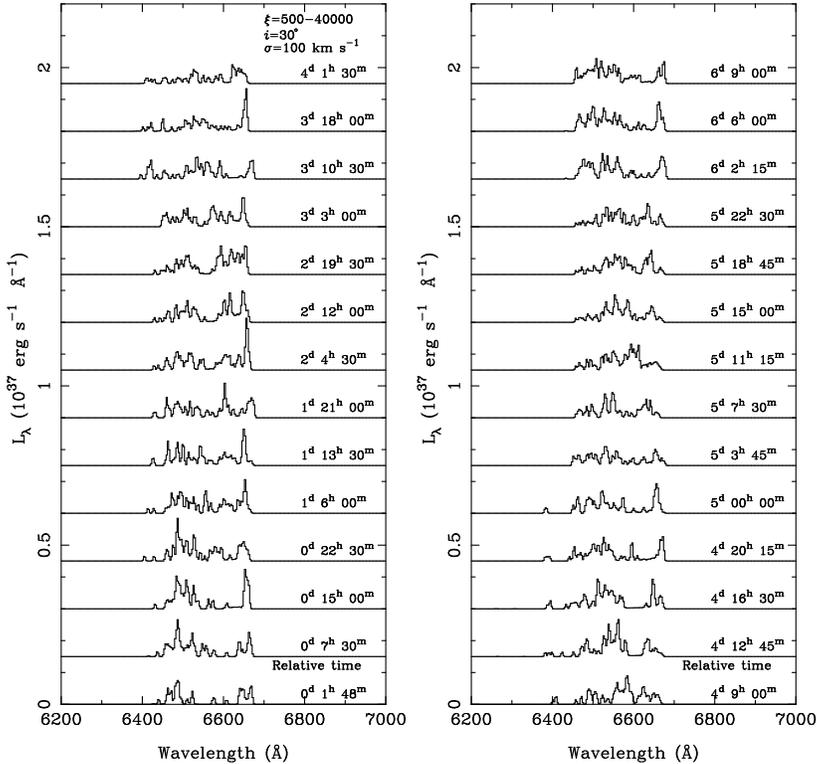}
  \caption{Sequence of H$\alpha$ profiles emitted from the
debris region $\xi\in(500,40\,000)$ (in units of $MG/c^2$) over period
of 6 days. The relative time from the beginning of the accretion phase
onto the black hole is marked next to each profile. The inclination of
the debris plane and the velocity shear are as marked on the
figure.\label{fig_1}}
\end{figure}

{Our model predicts prompt evolution of the profile shapes on the 
time scales of hours to days, as shown in Figure~\ref{fig_1}. The line
profiles can take a variety of shapes for different orientations of
the debris relative to the observer. Due to the very diverse
morphology of the debris, it is almost impossible to uniquely match
the multi-peaked profile with the exact emission geometry.
Nevertheless, the profile widths and shifts are strongly indicative of
the velocity distribution and the location of matter emitting the bulk
of the H$\alpha$ light. The profile shapes do not depend sensitively
on the shape of the UV/X-ray light curve illuminating the debris. They
strongly depend on the distance of the emitting material from the
central ionizing source, which is a consequence of the finite
propagation time of the ionization front and the redistribution of the
debris in phase space. If X-ray flares and the predicted variable
profiles could be observed from the same object they could be used to
identify the tidal disruption event in its early phase.}

\acknowledgements{We acknowledge the support of the Center for
Gravitational Wave Physics funded by the NSF under cooperative
agreement PHY-0114375, NSF grants PHY-9800973 and
PHY-0244788. T.B. also acknowledges the IAU support and the Zaccheus
Daniel Fellowship.}

\end{document}